\documentclass[a4paper, 11pt, dvips]{article}

\usepackage[latin1]{inputenc}
\usepackage{times}
\usepackage[T1]{fontenc}

\usepackage[top=2cm, bottom=2cm, left=2cm, right=2cm]{geometry}
\usepackage[normalem]{ulem}
\usepackage{sectsty}

\usepackage[dvips]{color}
\definecolor{linkcol}{rgb}{0,0,0.4} 
\definecolor{citecol}{rgb}{0.5,0,0}
 
\usepackage[a4paper, dvips]{hyperref}

\hypersetup
{
  bookmarks=true,
  bookmarksopen=false,
  breaklinks=true,
  hyperindex=true,
  pdfmenubar=true, %menubar shown
  pdfhighlight=/O, %effect of clicking on a link
  colorlinks=true, %couleurs sur les liens hypertextes
  pdfpagemode=None, %aucun mode de page
  pdfpagelayout=SinglePage, %ouverture en simple page
  pdffitwindow=true, %pages ouvertes entierement dans toute la fenetre
  linkcolor=linkcol, %couleur des liens hypertextes internes
  citecolor=citecol, %couleur des liens pour les citations
  urlcolor=linkcol %couleur des liens pour les url
}

\hypersetup%
{%
  pdftitle = {Bayesian Subset Simulation: a kriging-based subset
    simulation algorithm for the estimation of small probabilities of
    failure},%
  pdfauthor = {Ling Li, Julien Bect and Emmanuel Vazquez},%
  pdfsubject = {},%
  pdfkeywords = {Computer experiments; Sequential design; Subset
    Simulation; Probability of failure}%
}

\usepackage{amsmath, amsthm, amsfonts,amssymb,bm,bbm}

\usepackage{dsfont}             % \mathds
\renewcommand{\mathbb}{\mathds} % \mathds --> \mathbb

\newlength{\dhatheight}
\newcommand{\doublehat}[1]{%
    \settoheight{\dhatheight}{\ensuremath{\hat{#1}}}%
    \addtolength{\dhatheight}{-0.35ex}%
    \hat{\vphantom{\rule{1pt}{\dhatheight}}%
    \smash{\mskip -1mu\hat{#1}}}}

\usepackage{graphicx,psfrag,epsfig,subfigure}
\graphicspath{{./figs/}}
\usepackage{multirow,enumerate}

\usepackage{color}

\newcommand   \RR     {\mathbb{R}}

\newcommand   \XX     {\mathbb{X}}
\newcommand   \YY[1]  {\mathbb{Y}_{#1}} % #1=stage number 
\newcommand   \Y[2]   {Y_{#1}^{#2}}     % #1=stage number, #2=component number
\newcommand   \Yt[2]  {\widetilde{Y}_{#1,#2}} % same thing
\renewcommand \P      {\mathsf{P}}   % probability measure
\newcommand   \PX     {\P_{\XX}}     % probability measure on \XX
\newcommand   \E     {\mathsf{E}}   % expectation wrt \P
\newcommand   \Fcal   {\mathcal{F}}  % filtration
\newcommand   \Ical   {\mathcal{I}}  % "information": (X_1,\xi(X_1), ...)
\newcommand   \Ncal   {\mathcal{N}}  % Gaussian distribution
\newcommand   \ah[1]  {\hat{\alpha}_{#1}}        % Bayesian estimator (stage #1)
\newcommand   \ahh[1] {\doublehat{\alpha}_{#1}}  % MC approx. of \ah[1]
\renewcommand \u[1]   {u_{#1}}                   % threshold at stage #1
\newcommand   \one    {{\mathbb{1}}}

\newcommand   \dotvar {\,\bm{\cdot}\,}
\newcommand   \ddiff  {\mathrm{d}}

\newcommand   \dx     {\ddiff x}

\newcommand   \dPX    {\ddiff\PX}

\DeclareMathOperator \cov  {cov}

\renewcommand{\hat}{\widehat}

\newcommand \normPDF \varphi
\newcommand \normCDF \Phi

\newtheorem{remark}{Remark}
\makeatletter
\def\@seccntformat#1{\csname the#1\endcsname.\quad}
\sectionfont{\large\MakeUppercase}
\subsubsectionfont{\normalfont\uline}
\makeatother

\begin{document}
\setlength{\parindent}{0mm}
\setlength{\parskip}{3mm plus0.5mm minus0.5mm}
{
  \centering
  \Large{\textbf
    {
      Bayesian Subset Simulation: a kriging-based subset simulation algorithm\\
      for the estimation of small probabilities of failure
    }
  } \\
  \vspace{8mm}
  \large{
    \textbf{
      Ling Li$^{*}$, 	% Mark the contact author with asterisk *
      Julien Bect, 
      Emmanuel Vazquez
    }
  }\\[1em]
  SUPELEC Systems Sciences (E3S)\\
  Signal Processing and Electronic Systems Department\\
  Gif-sur-Yvette, France \\[1em]
}

\vspace{4mm}
\rule{\textwidth}{0.005in}
\textbf{Abstract:} 
%
% Write your abstract between these lines:
%---------------------------------------------------------------------------------------------------------------------------------------------------------------------
The estimation of small probabilities of failure from computer
simulations is a classical problem in engineering, and the Subset
Simulation algorithm proposed by Au \& Beck (Prob. Eng. Mech., 2001)
has become one of the most popular method to solve it. Subset
simulation has been shown to provide significant savings in the number
of simulations to achieve a given accuracy of estimation, with respect
to many other Monte Carlo approaches. The number of simulations
remains still quite high however, and this method can be impractical
for applications where an expensive-to-evaluate computer model is
involved.

We propose a new algorithm, called Bayesian Subset Simulation, that
takes the best from the Subset Simulation algorithm and from
sequential Bayesian methods based on kriging (also known as Gaussian
process modeling). The performance of this new algorithm is
illustrated using a test case from the literature. We are able to
report promising results. In addition, we provide a numerical study of
the statistical properties of the estimator.
%---------------------------------------------------------------------------------------------------------------------------------------------------------------------

\textbf{Keywords:}
%
% Write the keywords between these lines:
%---------------------------------------------------------------------------------------------------------------------------------------------------------------------
Computer experiments, Sequential design, Subset Simulation, Probability of failure
%---------------------------------------------------------------------------------------------------------------------------------------------------------------------
%
\vspace{-2mm}
\\
\rule{\textwidth}{0.005in}
%
%
%
% The text starts here:
%---------------------------------------------------------------------------------------------------------------------------------------------------------------------
\section{Introduction}

In this paper, we propose an algorithm called \textit{Bayesian Subset
  Simulation}~(BSS), that combines the Bayesian decision-theoretic framework
from our previous studies \cite{vaz:09:sysid, bect2010sequential} with the Subset
Simulation algorithm \cite{au01:_estim}.

Let $\Gamma = \{x \in \XX : f(x) > \u{} \}$ denote the excursion set of
a function $f:\XX\to\RR$ above a threshold $\u{} \in \RR$. We are
interested in estimating the probability $\alpha( \u{} ) := \PX
(\Gamma)$, which corresponds to the probability of failure of a
system for which $f$ is a function of performance~(see, e.g., \cite{bect2010sequential}).  If the probability~$\alpha(
\u{} )$ is small, estimating it using the Monte Carlo estimator
$\hat\alpha_m^{\tiny \rm MC} = 1/m\, \sum_{i=1}^m \one_{f(X_i) >u}$,
$X_i \stackrel{\scriptsize \rm i.i.d}{\sim} \PX$,
requires a large number of evaluations of $f$. If the
performance function $f$ is expensive to evaluate, this leads to use a
large amount of computational resources, and in some cases, it may be
even impossible to proceed in reasonable time. Estimating
small probabilities of failure with moderate computational resources
is a challenging topic. 

When $\alpha( \u{} )$ is small, the main problem with the estimator
$\hat\alpha_m^{\tiny \rm MC}$ is that the sample size $m$ must be large
in order to get a reasonably high probability of observing at least a
few samples in $\Gamma$.  In the literature, importance sampling
methods have been considered to generate more samples in the failure
region $\Gamma$. However, the success of this kind of methods relies
greatly on prior knowledge about the failure region $\Gamma$ and on a
relevant choice for the proposal sampling distribution.

The idea of Subset Simulation is to decompose the difficult problem of
generating samples in the failure region into a series of easier
problems, by introducing intermediate failure events. Let
$\u{0} = -\infty < \u{1} < \u{2} < \ldots < \u{T}
= \u{}$ be a sequence of increasing thresholds and define a
corresponding sequence of decreasing excursion sets $\Gamma_0:=\XX
\supseteq \Gamma_1 \supseteq \cdots \supseteq \Gamma_T:=\Gamma$, where
$\Gamma_{t} := \{x \in \XX: f(x) > \u{t} \}$, $ t = 1, \ldots,
T$.  Notice that $\Gamma_{t} = \bigcap_{i=1}^{t}\Gamma_{i}$. Then,
using the properties
\begin{equation}
  \label{equ:recurr-subsim}
  \begin{cases}
  \alpha(\u{0})   \;=\; 1\,, & \\
  \alpha(\u{t+1}) \;=\; \alpha(\u{t})\, \PX(\Gamma_{t+1}|\Gamma_{t})\,, &
  t \geq 0\,,
  \end{cases}
\end{equation}
$\alpha( \u{} )$ can be rewritten as a product of conditional probabilities:
\begin{equation}
  \label{eq:subset-sampling}
  \alpha( \u{} ) \;=\; \PX\bigl( \Gamma_{T} \bigr)
  \;=\; \prod_{t=0}^{T-1}\PX(\Gamma_{t+1}|\Gamma_{t})\,.
\end{equation}
Thus, the idea of Subset Simulation is to replace the problem of estimating
the small probability $\alpha( \u{} )$ by that of estimating the \emph{higher}
conditional probabilities $\PX(\Gamma_{t+1}|\Gamma_{t})$, $0 \le
t < T$.

In \cite{au01:_estim}, a standard Monte Carlo simulation method is
used to estimate $\PX(\Gamma_{1}) = \PX(\Gamma_{1}|\Gamma_{0})$. For
the other conditional probabilities, a Markov Chain Monte Carlo method
is used to simulate samples in $\Gamma_{t}$, and then
$\PX(\Gamma_{t+1}|\Gamma_{t})$ is estimated using a Monte Carlo
method.  Due to the direct use of Monte Carlo method, the number of
evaluations needed remains still quite high. For many practical
applications where the performance function corresponds to an
expensive-to-evaluate computer model, it is not applicable. Note that
the Subset Simulation algorithm has recently caught the attention of
the Sequential Monte Carlo (SMC) community: using standard tools from
the SMC literature, \cite{cerou:2011:smc-rare-events} derives several
theoretical results about some very close versions of the Subset
Sampling algorithm.

In this work, we propose an algorithm that takes advantage of a
Gaussian process prior about $f$ in order to decrease the number of
evaluations needed to estimate the conditional probabilities
$\P_{\XX}(\Gamma_{t+1}|\Gamma_{t})$.  The Gaussian process model makes
it possible to assess the uncertainty about the position of the
intermediate excursion sets $\Gamma_t$, given a set of past evaluation
results. The idea has its roots in the field of design and analysis of
computer experiments (see, e.g., \cite{sac89, Currin91, Welch92,
  oakley:2002:bayesian, oakley:2004:probabilistic, oakley:2004:perc,
  bayarri:2007:fvcm}). More specifically, kriging-based sequential
strategies for the estimation of a probability of failure are closely
related to the field of Bayesian global optimization \cite{Mockus,
  MockusB, Jones, villemonteix:2008:phd, villemonteix:2009:iago,
  GinsbPhd}.

The paper is organized as follows. In Section \ref{sec:BSS}, we give a
detailed presentation of our new Bayesian Subset Simulation
algorithm. In Section~\ref{sec:num}, we apply the algorithm on an
example from the literature, and we perform numerical simulations to
investigate the performance of the proposed algorithm. A comparison with
Subset Simulation and classical Monte Carlo methods is provided. Finally, we
conclude in Section~\ref{sec:discuss}.

\textbf{Remark.} An algorithm involving kriging-based adaptive
sampling and Subset Simulation has been recently proposed by
V. Dubourg and co-authors \cite{dubourg:2011:rbdo, Dubourg:2011:phd}
to address the problem of Reliability-Based Design Optimization
(RBDO). Their approach is different from the one proposed in
this paper, which addresses the problem of reliability analysis.

\section{Bayesian Subset Simulation algorithm}
\label{sec:BSS}

\subsection{Algorithm}
\label{sec:optim_strategies}

Our objective is to build an estimator of~$\alpha( \u{T} )$ from the
evaluations results of~$f$ at a number of points $X_1, X_2, \ldots,
X_N \in\XX$. Let $\xi$ be a random process modeling our prior
knowledge about $f$, and for each $n\geq 0$, denote by $\Fcal_n$ the
$\sigma$-algebra generated by $X_1, \xi(X_1), \ldots, X_n,
\xi(X_n)$. A natural
Bayesian estimator of $\alpha(\u{t})$ using $n_t$ evaluations is the
posterior mean
\begin{equation}
  \label{eq:estimator1}
  \ah{t} \;=\; \E_{n_t}\left( \alpha(\u{t}) \right)
  \;=\; \E_{n_t} \left( \int_{\XX} \one_{\xi  > \u{t}}\, \dPX \right) 
  \;=\; \int_\XX g_t\, \dPX,
\end{equation}
where $g_t:x\in\XX \mapsto \P_{n_t} \bigl(\, \xi(x) > \u{t} \,\bigr)$
and $\E_n$ (resp. $\P_n$) denotes the conditional expectation
(resp. conditional probability) with respect to $\Fcal_n$~(see \cite{bect2010sequential}). Note that,
$g_{t}(x)$ can be readily computed for any $x$ using kriging~(see, e.g.,
\cite{bect2010sequential}).

Assume now that $\PX$ has a probability density function
$p_{\XX}$ and consider the sequence of probability density functions
$q_t$, $0 \le t \le T$, defined by
\begin{equation}
  \label{equ:qt}
  q_{t}(x) \;=\; \frac{1}{\hat{\alpha}_t}\, p_{\XX}(x)\, g_t(x).
\end{equation}
We can write a recurrence relation similar to~\eqref{equ:recurr-subsim}
for the sequence of Bayesian estimators~$\ah{t}$:
\begin{equation}
  \label{equ:recurr-bss}
  \ah{t} 
  \;=\; \int g_t(x)\, p_{\XX}(x)\, \dx
  \;=\; \ah{t-1} \int \frac{g_t(x)}{g_{t-1}(x)}\, q_{t-1}(x)\, \dx.
\end{equation}
The idea of our new algorithm, that we call Bayesian Subset Simulation,
is to construct recursively a Monte Carlo approximation~$\ahh{T}$ of the
Bayesian estimator~$\ah{t}$, using~\eqref{equ:recurr-bss} and sequential
Monte Carlo simulation (SMC) (see,
e.g., \cite{delmoral:2006:sequential}) for the evaluation of the integral
with respect to~$q_{t-1}$ on the right-hand side. More precisely,
denoting by~$m$ the size of the Monte Carlo sample, we will use the
recurrence relation
\begin{equation}
\label{eq:discre}
  \ahh{t} = \ahh{t-1} \times \frac{1}{m} \sum_{i=1}^{m}
  \frac{g_{t}(\Y{t-1}{i})}{g_{t-1}(\Y{t-1}{i})},
  \quad 1 \le t \le T,
\end{equation}
where variables $\Y{t-1}{1}, \ldots, \Y{t-1}{m}$ are distributed
according to\footnote{By ``distributed according to'', it is not meant
  that $\Y{t-1}{1}, \ldots, \Y{t-1}{m}$ are independent and
  identically distributed. This is never the case in sequential
  Monte-Carlo techniques. What we mean is that the sample $\Y{t-1}{1},
  \ldots, \Y{t-1}{m}$ is \textit{targetting} the density~$q_{t-1}$ in the
  sense of, e.g., \cite{douc:2008:limit}.} the density $q_{t-1}$,
which leads to
\begin{equation}
  \label{eq:bss-estimator}
  \ahh{T} \;=\;
  \prod_{t=0}^{T-1} \frac{1}{m} \sum_{i=1}^{m}
\frac{g_{t+1}(\Y{t}{i})}{g_{t}(\Y{t}{i})}\,.
\end{equation}

The connection between the proposed algorithm and the original Subset
Simulation algorithm is clear from the similarity between
the recurrence relations~\eqref{equ:recurr-subsim}
and~\eqref{equ:recurr-bss}, and the use of SMC simulation in both
algorithms to construct recursively a ``product-type'' estimator of
the probability of failure (see also in \cite{delmoral:2006:sequential},
Section 3.2.1, where this
type of estimator is mentioned in a very general SMC
framework). 

% \textcolor{red}{\cite{au01:_estim} proposed a modified
%   Metropolis-Hasting method in order to increase the ratio of acceptance
% of the particles. In our proposed algorithm, a multi-chain Gibbs
% sampling method, which is equivalent to the composition of $p$
% Metropolis-Hasting method, is used. The particles obtained at the last
% of the chain is the generated sample.}

Our choice for the sequence of densities $q_1, \ldots, q_T$ also
relates to the original Subset Simulation algorithm. Indeed, note that
$q_t(x) \propto \E_{n_t} \bigl( \one_{\xi > \u{t}}\, p_{\XX} \bigr)$, and recall
that $\tilde q_t \propto \one_{\xi > \u{t}}\, p_{\XX}$ is the
distribution used in the Subset Simulation algorithm at
stage~$t$. (This choice of instrumental density is also used by
\cite{dubourg:mbis,dubourg:icasp11} in the context of a two-stage
kriging-based adaptive importance sampling algorithm. This is indeed a
quite natural choice, since $\tilde q_T \propto \one_{\xi > u}\, p_{\XX}$ is
the optimal instrumental density for the estimation of $\alpha( \u{} )$ by
importance sampling.)

\subsection{Implementation}
\label{sec:implemen}

This section gives implementation details for our Bayesian Subset
Simulation algorithm, the principle of which has been described in the
Section~\ref{sec:optim_strategies}. The pseudo-code for the algorithm
is presented in Table~\ref{tab:imple}.

The initial Monte Carlo sample $\YY{0} = \{ \Y{0}{1}, \ldots, \Y{0}{m}
\}$ is a set of independent random variables drawn from the
density~$q_0 = p_\XX$---in other words, we start with a classical
Monte Carlo simulation step. At each subsequent stage~$t \ge 1$, a new
sample~$\YY{t}$ is produced from the previous one using the basic
reweight/resample/move steps of SMC simulation (see \cite{delmoral:2006:sequential} and the
references therein). In this article,
resampling is carried out using a multinomial sampling scheme, and the
move step relies on a fixed-scan Metropolis-within-Gibbs algorithm
as in \cite{au01:_estim} with a Gaussian-random-walk
proposal distribution for each coordinate (for more
information on these techniques, see, e.g., \cite{robert:2004:monte}).

A number~$N_t$ of evaluations of the performance function is
done at each stage of the algorithm. This number is meant to be much
smaller than the size~$m$ of the Monte Carlo sample---which would be
the number of evaluations in the classical Subset Sampling
algorithm. For the initialization stage ($t = 0$), we choose a space
filling set of points~$\YY{0}$ as usual in the design of computer
experiments \cite{santner:2003:dace}. At each subsequent stage, we
use $N_t$ iteration of a SUR sampling strategy
\cite{bect2010sequential} targeting the threshold~$\u{t}$ to select the
evaluation points. Adaptive techniques to choose the sequence of
thresholds and the number of points per stage are presented in the
following sections.

\begin{remark}
  The resampling step could most certainly benefit from more elaborate
  schemes, such as the residual resampling scheme
  \cite{hol:2006:resampling, douc2005comparison, bolic2003new}. The
  comparison of resampling schemes is left for future work.
\end{remark}

\begin{table}[t]% [htbp]
  \caption{Algorithm of Bayesian Subset Simulation}
  \label{tab:imple}
  \noindent\ignorespaces%
  \rule{\textwidth}{.2pt}%
  \par
  \begin{enumerate}[a)]
  \item Initialize~(Stage $0$): 
    \begin{description}
    \item[1.] Generate a MC sample $\YY{t} = \{ \Y{0}{1}, \ldots, \Y{0}{m}
      \}$, drawn according to the distribution $\PX$
    \item[2.] Initial DoE $\Ical_{n} = \{(X_1,
      f(X_1)),\ldots,(X_{n_{0}},f(X_{n_{0}}))\}$~(maximin)
    \item[3.] Choose kriging model, estimate parameters $k_{\theta}$
    \end{description}
    
    \medbreak
    
  \item At each stage $t (t = 1 \ldots T)$:
    \begin{description}
    \item[1.] Compute the kriging predictor $\hat{f}_{n}^{t-1}$, and
      choose threshold $\tilde{u}^{t-1}$
    \item[2.] Select and evaluate $N_t$ new points using a SUR
      sampling criterion for the threshold $\u{t}$.
    \item[3.] Update $\Ical_{n}$,
      adjust intermediate threshold $\u{t-1}$ according to $\hat{f}_{n}^{t-1}$
    \item[4.] Regenerate a new sample $\YY{t}$:
      \begin{flushleft}
        \begin{description}
        \item[4.1] \textbf{reweight}: calculate weights: $w_{i}^{t} \propto
          g_{t}(\Y{t-1}{i}) / g_{t-1}(\Y{t-1}{i})$
        \item[4.2] \textbf{resample}: generate a sample
          $\Yt{t-1}{i}$ according to weights
        \item[4.3] \textbf{move}: for each $i \le m$, $\Y{t}{i}
          \backsim K\bigl( \Yt{t-1}{i}, \dotvar \bigr)$
        \end{description}
      \end{flushleft}
    \end{description}

    \medbreak

  \item The final probability of failure is calculated by 
    $$ \hat{\alpha} = \prod_{t=0}^{T-1} \Big( \frac{1}{m} \sum_{i=1}^{m}
    \frac{g_{t+1}(\Y{t}{i})}{g_{t}(\Y{t}{i})}  \Big)     $$
    
  \end{enumerate}
  \noindent\ignorespaces%
  \rule{\textwidth}{.2pt}%
\end{table}
 
\subsection{Adaptive choice of the thresholds $u_{t}$}

It can be proved that, for an idealized\footnote{assuming that $\Y{t}{1}$,
  \ldots, $\Y{t}{m}$ are independent and identically distributed
  according to~$q_t$.} Subset Simulation algorithm with fixed
thresholds $\u{0} < \u{1} < \cdots < \u{T} = \u{}$, it is optimal to
choose the thresholds to make all conditional probabilities
$\PX\bigl( \Gamma_{t+1} | \Gamma_{t} \bigr)$ equal to some constant
value~(see \cite{cerou:2011:smc-rare-events}, Section 2.4). This leads to the
idea of choosing the thresholds adaptively in such a way that, in the
product estimate
\begin{equation*}
  \ah{T}^{\mathrm{SubSamp}} \;=\;
  \prod_{t=1}^T \frac{1}{m} \sum_{i=1}^m
  \one_{\Gamma_t}\bigl( \Y{t-1}{i} \bigr),
\end{equation*}
each term but the last is equal to some predefined constant~$p_0$. In
other words, $\u{t}$ is chosen as the $(1-p_0)$-quantile
of~$\YY{t-1}$. This idea was first suggested by \cite{au01:_estim} in Section
5.2, on the heuristic ground that the algorithm
should perform well when the conditional probabilities are neither too
small (otherwise they are hard to estimate) nor too large (otherwise a
large number of stages is required). The asymptotic behavior of the
resulting algorithm, when $m$ is large, has been analyzed by
\cite{cerou:2011:smc-rare-events}.

In Bayesian Subset Simulation, we propose to choose the thresholds
adaptively using a similar approach. More precisely, considering the
product form of the estimator~\eqref{eq:bss-estimator}, we suggest to
choose~$\u{t}$ in such a way that
\begin{equation*}
  \frac{1}{m} \sum_{i=1}^{m} \frac{%
    g_{t+1}(\Y{t}{i})%
  }{%
    g_{t}(\Y{t}{i})%
  } 
  \;=\; p_{0}.
\label{eq:p0}
\end{equation*}
The equation can be easily solved since the left-hand side is a strictly decreasing
function of~$\u{t}$. 

\begin{remark}
Note that \cite{cerou:2011:smc-rare-events} proved that choosing adaptive levels in
Subset Simulation
introduces a positive bias of order $1/m$, which is
negligible compared to its standard deviation. 
\end{remark}

\subsection{Adaptive choice of the number~$N_t$ of evaluation at each stage}
\label{subsec:autoSUR}

In this section, we propose a technique to choose adaptively the
number~$N_t$ of evaluations of the performance function that must be
done at each stage of the algorithm. 

Let us assume that~$t \ge 1$ is the current stage number; at the
beginning of the stage, $n_{t-1}$ evaluations of the performance
function are available from previous stages. After several additional
evaluations, the number of available observations of~$f$ is $n \ge
n_{t-1}$. Then, for each $i \in \{1, \ldots, m \}$, the probability of
misclassification\footnote{See \cite{bect2010sequential} Section
  2.4 for more information} of~$x \in \XX$ with
respect to the threshold~$\u{t}$ is
\begin{equation*}
  \tau_{t,n} (x) \;=\; \min \Bigl(%
  p_{n}( x, \u{t} ),\,
  1 - p_{n}(x,, \u{t} )%
  \Bigr),
\end{equation*}
where $p_n(x,u) = \E_n\left( \one_{\xi(x) > u } \right)$~(see \cite{bect2010sequential}). We shall
decide to stop adding new evaluations at stage $t$ when
\begin{equation*}
  \frac{1}{m}\,
  \sum_{i=1}^m \tau_{t,n}\left( \Y{t-1}{i} \right) \le \eta,
\end{equation*}
for some prescribed~$\eta > 0$.

\section{Numerical results}
\label{sec:num}

In this section, we apply the proposed algorithm on a simple 2D test
case from the structural reliability literature. The problem under
consideration is the deviation of a cantilever beam, with a
rectangular cross-section, and subjected to a uniform load
\cite{gayton2003cq,rajashekhar1993new}.  The performance function is:
\begin{equation}
f(x_{1}, x_{2}) = 18.46154 - 7.476923 \times 10^{10}\, \frac{x_{1}}{{x_{2}}^{3}}\,.
\end{equation}  

\begin{figure}
  \centering
  \includegraphics[width=.75\textwidth]{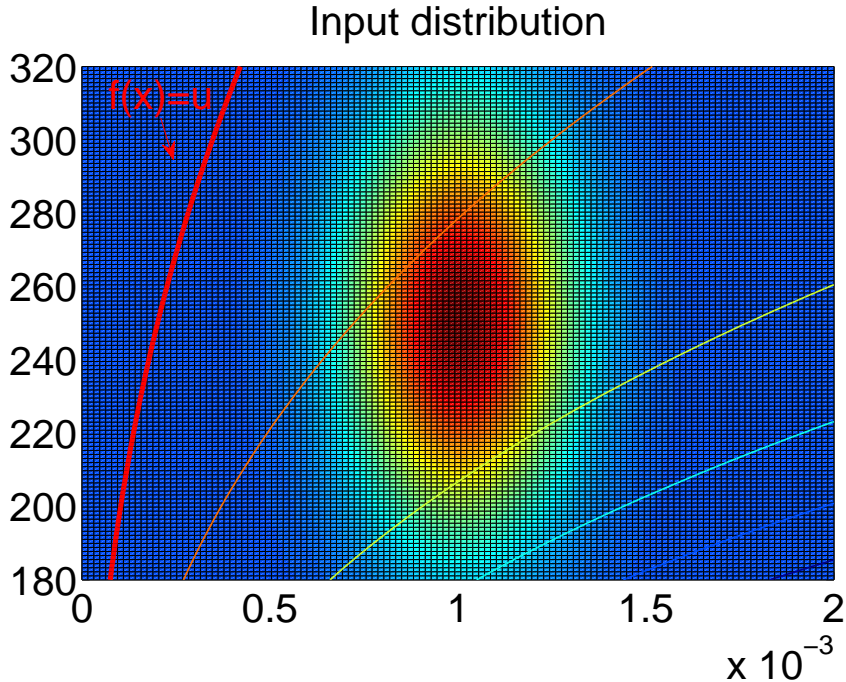}
  \vspace{-0.5em}
  \caption{Input distribution and contour plot of
    the performance function}
  \label{fig:ex3:dis}
\end{figure}

The uncertain factors are $x_{1}$ and $x_{2}$, which are supposed to
be independent and normally distributed, as specified in
Table~\ref{tab:var2}. We use $\u{} = 17.8$ as the threshold for the
definition of the failure event. The probability of failure, which
will be used as reference estimator, obtained using
$\hat\alpha^{\tiny\rm MC}_m$ with $m = 10^{8}$, is approximately $3.85
\times 10^{-5}$ (with a coefficient variance of about $1 / \sqrt{m\,
  \alpha} \approx 1.61\%$). Figure~\ref{fig:ex3:dis} shows the
distribution of the input factors along with a contour plot of
$f$. Notice that the failure region is quite far from the center
region of the input distribution.

\begin{table}
  \caption{Random input factors}
  \centering
  \begin{tabular}{c c c c}
    \hline
    Variable & Distribution & Mean $m$ & Standard deviation $\sigma$ \\\hline
    $x_{1}$ & $\Ncal$ & $0.001$ & $0.0002$  \\  
    $x_{2}$ & $\Ncal$ & $250$ & $37.5$  \\\hline
  \end{tabular}
  \label{tab:var2}
  \vspace{1em}
\end{table}

In the Bayesian Subset Simulation algorithm, we set an initial design of
size $N_0=10$ which is equal to five times the
dimension $d$ of the input space~(In the literature, very little is known
about the problem of choosing $N_{0}$, however some authors recommend to
start with a sample size proportional to the dimension $d$, see~
\cite{loeppky:2009:css}). Concerning the choice of $N_{0}$, we decide
to apply a greedy MAXMIN algorithm and sequentially choose the points
which will maximize the minimal Euclidean distance between any two
points in the initial Monte Carlo
sample $\YY{0}$.  At each stage, we choose a Monte Carlo
sample of size $m=1000$. A Gaussian process with constant
unknown mean and a Mat\'ern covariance function is used as our prior
information about $f$. The parameters of the Mat\'ern covariance
functions are estimated on the initial design by REML~(see, e.g.,
\cite{Ste99}). In this experiment, we follow the common practice of
re-estimating the parameters of the covariance function during the
sequential strategy, and update the covariance function after each evaluation. The target conditional probability between successive
thesholds is set to $p_{0} = 0.1$. The intermediate threshold $\u{t}$
is chosen by the criterion \eqref{eq:p0}. The proposal distribution $q_{t}$ for a Gibbs sampling is a
Gaussian distribution  $\Ncal(0, \sigma^2)$, where
$\sigma^2$ is specified in Table~\ref{tab:var2}. The stopping criterion for
the adaptive SUR strategy is set to $\eta_{t} = 10^{-6}$ (for $t = 1,
\ldots, T-1$) and $\eta_{T} = 10^{-7}$. 

\begin{figure}[ht!]
  \psfrag{x1}{\tiny $x_{1}$}
  \psfrag{x2}{\tiny $x_{2}$}
  \psfrag{ut}{\tiny $u_{t}, t=1,\ldots, T$}
  \centering
  \subfigure{\label{ex3:doe:t1} 
    \includegraphics[width = .45\textwidth]{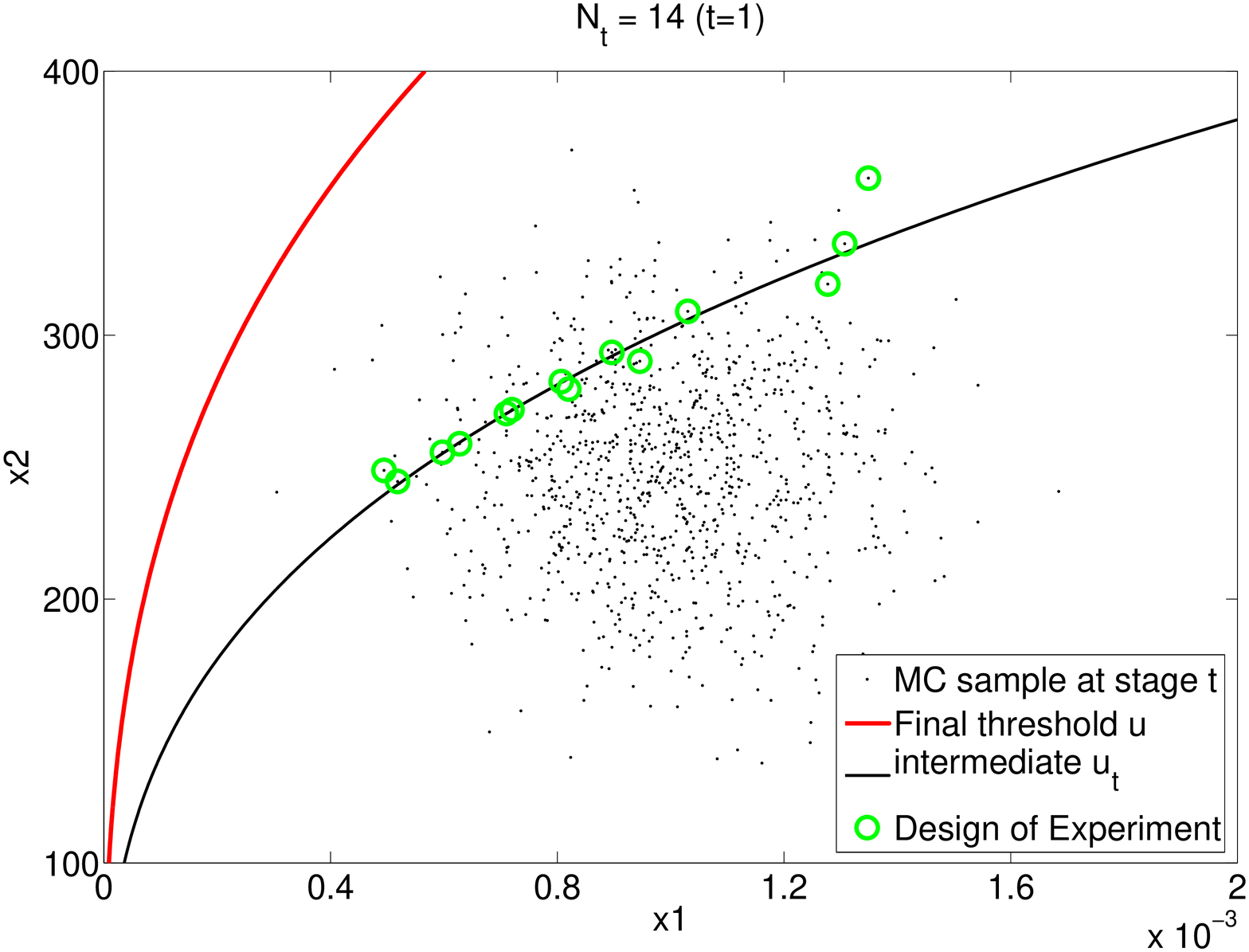}}
  \subfigure{\label{ex3:doe:t2}
    \includegraphics[width =.45\textwidth]{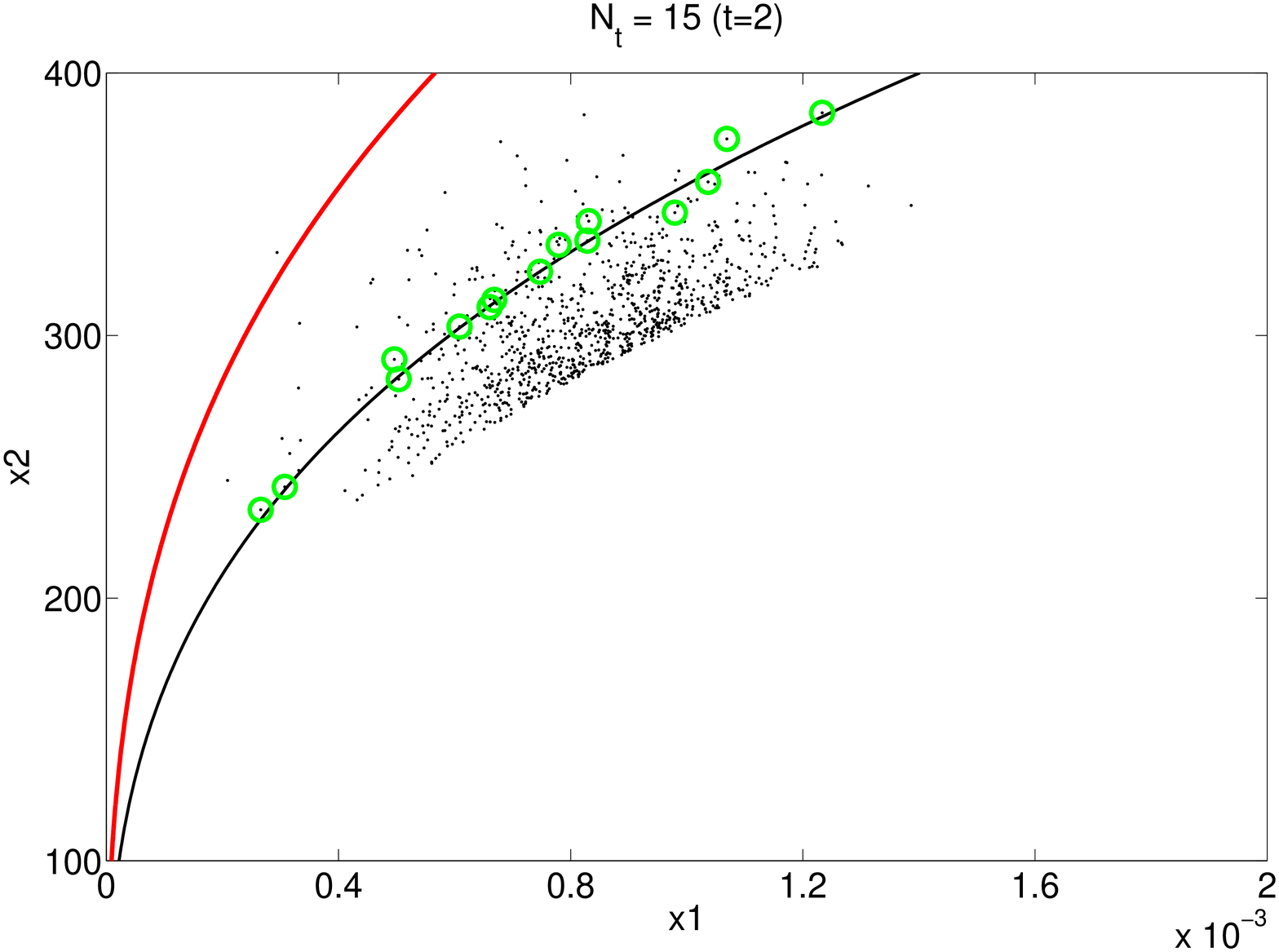}} 
  \\
  \subfigure{\label{ex3:doe:t3}
    \includegraphics[width = .45\textwidth]{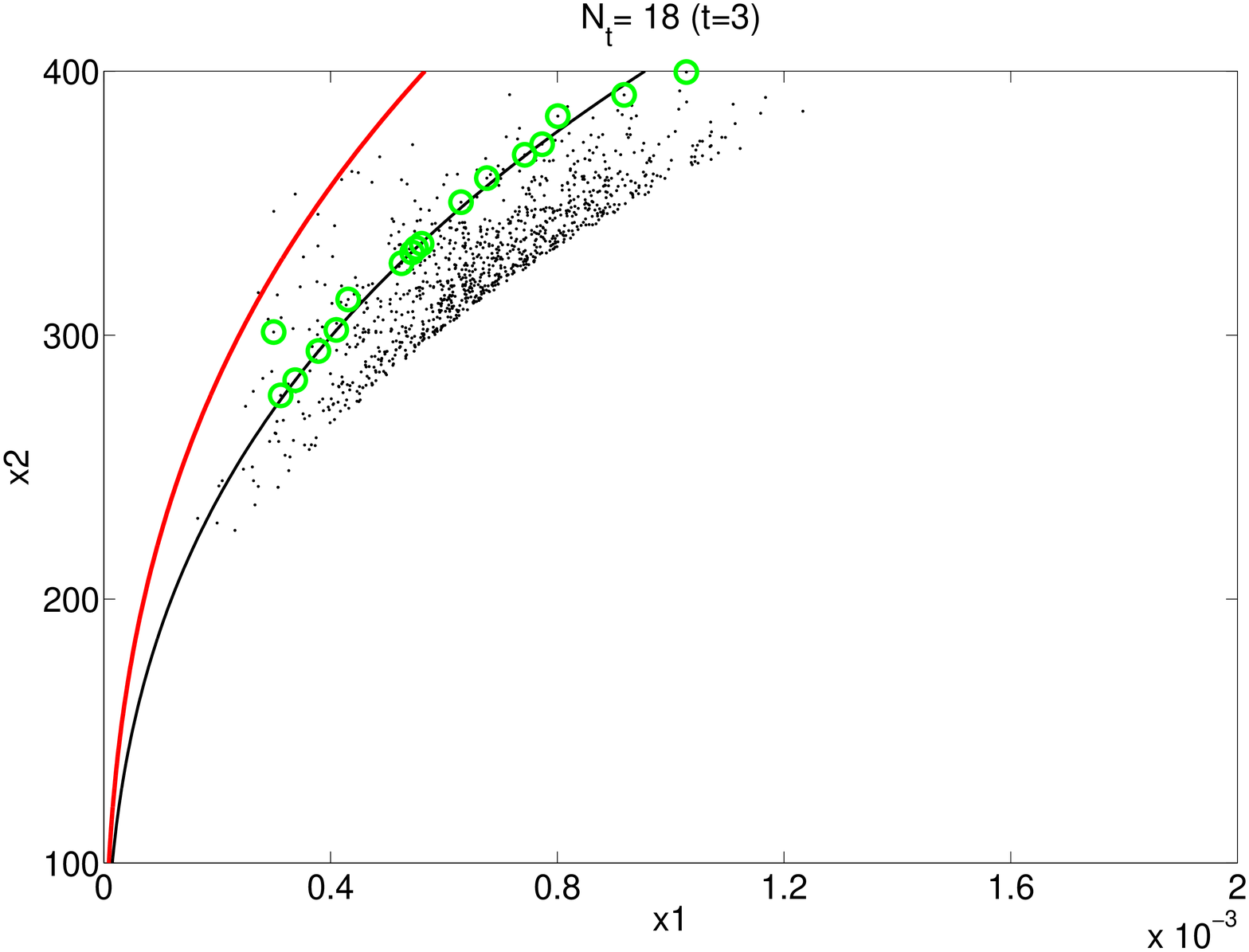}}
  \subfigure{\label{ex3:doe:FINA}
    \includegraphics[width = .45\textwidth]{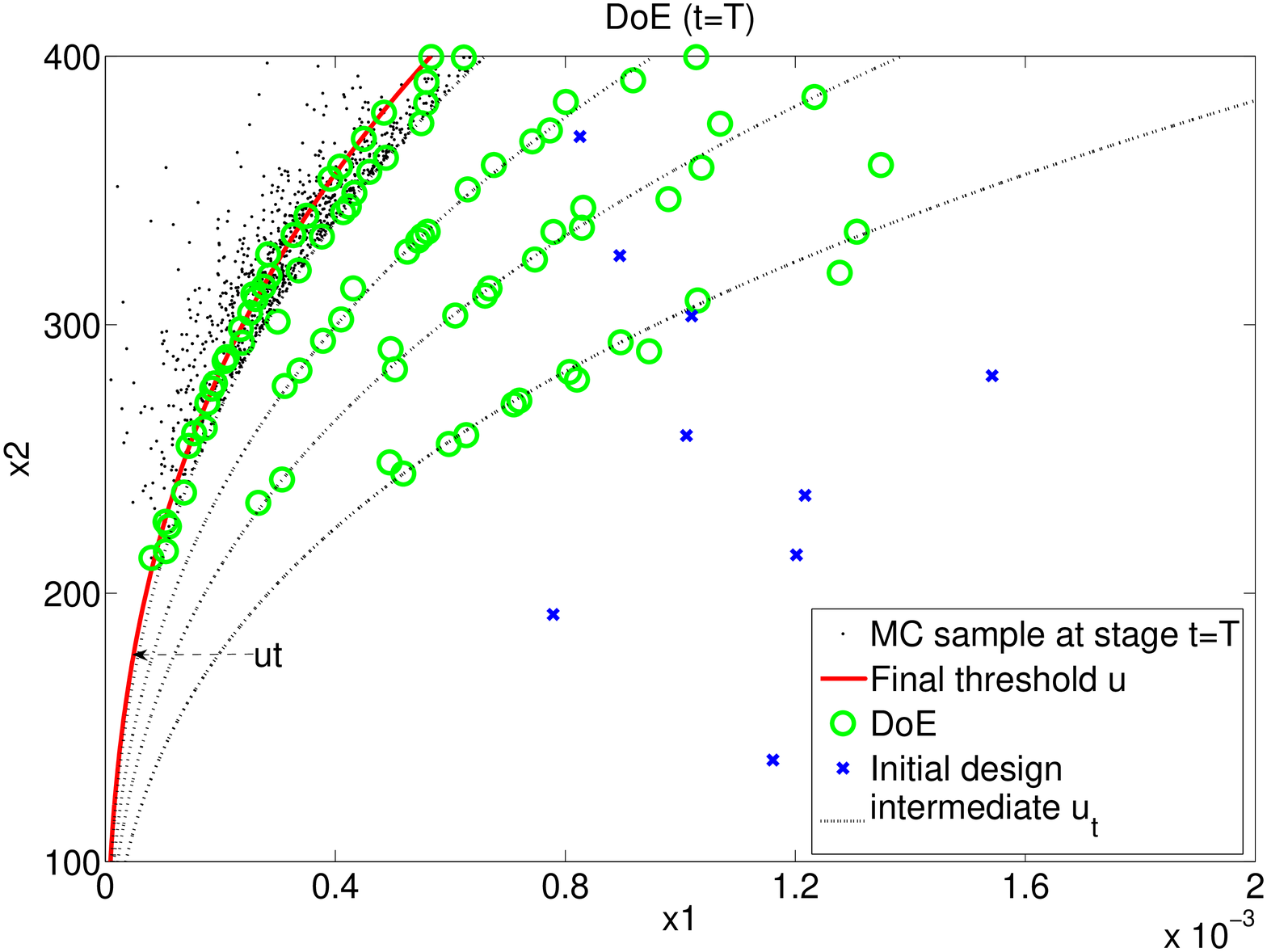}}
  \caption{Evaluations selected by the Bayesian Subset Simulation strategy at stage
    $t=1,2,3$ and the final Design of Experiment~(DoE). }
  \label{fig:ex3:doe}
  \vspace{2em}
\end{figure}

Figure~\ref{fig:ex3:doe} shows
the Design of Experiment~(DoE) selected by the algorithm
at stage $t=1,2,3$ and the last stage for one run.  Table \ref{tab:ex2:no} lists
the number of evaluations~(rounded to integer) at
each stage averaged over $50$ runs. We can see that an average total of evaluations $N =
\sum_{i=0}^{T}N_{t} = 104$ are
needed for our proposed Bayesian Subset Simulation, while for Subset Simulation,
the number is $1000 + 900 \times 4 = 4600$.
\begin{table}
  \caption{Average number of evaluations at each stage.} 
  \label{tab:ex2:no}
  \centering
  \begin{tabular}{c|c|c|c|c|c}
    \hline
    $N_{t}$  & $1$ & $2$  & $3$ &$4$ &$5$
    \\ \hline
    Sub-Sim &  $1000$ & $900$ & $900$ & $900$ & $900$ \\ \hline
    Bayesian Sub-Sim & $14$ & $17$ & $17$ & $18$ & $28$\\ \hline
  \end{tabular}
  \vspace{1em}
\end{table}

To evaluate the statistical properties of the estimator, we consider
the absolute relative bias
\begin{equation}
\label{eq:arb}
\kappa = \left|\frac{\E(\hat{\alpha}) - \alpha}{\alpha} \right|
\end{equation}
and the coefficient of variation
\begin{equation}
\label{eq:cov}
\cov = \frac{\delta(\hat{\alpha})}{\alpha},
\end{equation}
where $\E(\hat{\alpha})$ is the average mean and $\delta(\hat{\alpha})$ is the standard deviation of the estimator $\hat{\alpha}$.

Table~\ref{tab:ex2:comp} shows the results of the comparison of our proposed
Bayesian Subset Sampling algorithm with the Subset Simulation algorithm in
\cite{au01:_estim}. Crude Monte Carlo sampling is used as the reference
probability of failure. For the fairness of the comparison, we
set the same intermediate probability $p_{0}=0.1$, and sample
size $m=1000$ in both methods. Fifty independent runs are performed to evaluate the average
of both methods. For Bayesian Subset
Simulation method, as $N$ is different for each run, we show the minimal
and maximal of $N$ from $50$ runs.

\begin{table}[ht!]
\caption{Compare with Monte Carlo approach and Subset Simulation}
  \label{tab:ex2:comp}
  \centering
 \begin{tabular}{c|c|c|c|c|c|c}
\hline
 Method &  $m$ &  $N$ & $\hat{\alpha}~(10^{-5})$& $\delta(\hat{\alpha})~(10^{-5})$  & $\kappa$ & $\cov$  \\ \hline
 MCS & $10^{8}$ & $10^{8}$ & $3.8500$ &  $0.062$   & $0$ & $1.6\%$ \\  \hline
 Sub-Sim & $1000$ & $4600$ & $3.9078$ &  $2.470$  & $1.5 \%$  & $63.2 \%$   \\ \hline
Bayesian Sub-Sim$$ & $1000$
& $[94, 109]$ & $3.7020$ & $0.618$  &  $4.4\%$ & $16.7\%$ \\  \hline
\end{tabular}
\end{table}

\section{Conclusion}
\label{sec:discuss}

In this paper, we propose a new algorithm called Bayesian Subset
Simulation for estimating small probabilities of failure in a context of
very expensive simulations. This algorithm combines the main ideas of
the Subset Simulation algorithm and the SUR strategies developed in our
recent work \cite{bect2010sequential}.

Our preliminary results show that the number of evaluations is
dramatically decreased compared to the original Subset Simulation algorithm, while
keeping a small bias and coefficient of variation.

Our future work will try to improve further the properties of our
algorithm regarding the bias and the variance of the estimator. We
shall also test and validate the approach  on more challenging examples.

\section*{Acknowledgments}

  The research of Ling Li, Julien Bect and Emmanuel Vazquez was partially
  funded by the French \emph{Fond Unique Interminist\'eriel} (FUI) in the
  context of the project \textsc{CSDL}.
%---------------------------------------------------------------------------------------------------------------------------------------------------------------------
% The text ends here. 
%
%
%
% The bibliography:
%---------------------------------------------------------------------------------------------------------------------------------------------------------------------
\bibliographystyle{unsrt}
\bibliography{Bibliography}

\begin{thebibliography}{10}

\bibitem{vaz:09:sysid}
E.~Vazquez and J.~Bect.
\newblock A sequential {B}ayesian algorithm to estimate a probability of
  failure.
\newblock In {\em Proceedings of the 15th {IFAC} {S}ymposium on {S}ystem
  {I}dentification, {SYSID} 2009 15th {IFAC} {S}ymposium on {S}ystem
  {I}dentification, {SYSID} 2009}, {S}aint-{M}alo {F}rance, 2009.

\bibitem{bect2010sequential}
J.~Bect, D.~Ginsbourger, L.~Li, V.~Picheny, and E.~Vazquez.
\newblock Sequential design of computer experiments for the estimation of a
  probability of failure.
\newblock {\em Statistics and Computing}, pages 1--21, 2010.

\bibitem{au01:_estim}
S.~K. Au and J.~Beck.
\newblock Estimation of small failure probabilities in high dimensions by
  subset simulation.
\newblock {\em Probab. Engrg. Mechan.}, 16(4):263--277, 2001.

\bibitem{cerou:2011:smc-rare-events}
F.~C{\'{e}}rou, P.~Del~Moral, T.~Furon, and A.~Guyader.
\newblock Sequential monte carlo for rare event estimation.
\newblock {\em Statistics and Computing}, pages 1--14, 2011.

\bibitem{sac89}
J.~Sacks, W.~J. Welch, T.~J. Mitchell, and H.~P. Wynn.
\newblock Design and analysis of computer experiments.
\newblock {\em Statistical Science}, 4(4):409--435, 1989.

\bibitem{Currin91}
C.~Currin, T.~Mitchell, M.~Morris, and D.~Ylvisaker.
\newblock Bayesian prediction of deterministic functions, with applications to
  the design and analysis of computer experiments.
\newblock {\em J. Amer. Statist. Assoc.}, 86(416):953--963, 1991.

\bibitem{Welch92}
W.~J. Welch, R.~J. Buck, J.~Sacks, H.~P. Wynn, T.~J. Mitchell, and M.~D.
  Morris.
\newblock Screening, predicting and computer experiments.
\newblock {\em Technometrics}, 34:15--25, 1992.

\bibitem{oakley:2002:bayesian}
J.~Oakley and A.~O'Hagan.
\newblock {Bayesian inference for the uncertainty distribution of computer
  model outputs}.
\newblock {\em Biometrika}, 89(4), 2002.

\bibitem{oakley:2004:probabilistic}
J.E. Oakley and A.~O'Hagan.
\newblock {Probabilistic sensitivity analysis of complex models: a Bayesian
  approach}.
\newblock {\em Journal of the Royal Statistical Society: Series B (Statistical
  Methodology)}, 66(3):751--769, 2004.

\bibitem{oakley:2004:perc}
J.~Oakley.
\newblock Estimating percentiles of uncertain computer code outputs.
\newblock {\em J. Roy. Statist. Soc. Ser. C}, 53(1):83--93, 2004.

\bibitem{bayarri:2007:fvcm}
M.~J. Bayarri, J.~O. Berger, R.~Paulo, J.~Sacks, J.~A. Cafeo, J.~Cavendish,
  C.-H. Lin, and J.~Tu.
\newblock A framework for validation of computer models.
\newblock {\em Technometrics}, 49(2):138--154, 2007.

\bibitem{Mockus}
J.~Mockus, V.~Tiesis, and A.~Zilinskas.
\newblock The application of {B}ayesian methods for seeking the extremum.
\newblock In L.~Dixon and Eds~G. Szego, editors, {\em Towards Global
  Optimization}, volume~2, pages 117--129. Elsevier, 1978.

\bibitem{MockusB}
J.~Mockus.
\newblock {\em Bayesian Approach to Global Optimization. Theory and
  Applications}.
\newblock Kluwer Academic Publisher, Dordrecht, 1989.

\bibitem{Jones}
D.~R. Jones, M.~Schonlau, and J.~William.
\newblock Efficient global optimization of expensive black-box functions.
\newblock {\em Journal of Global Optimization}, 13(4):455--492, 1998.

\bibitem{villemonteix:2008:phd}
J.~Villemonteix.
\newblock {\em Optimisation de fonctions co\^{u}teuses}.
\newblock PhD thesis, Universit\'{e} Paris-Sud XI, Facult\'{e} des Sciences
  d'Orsay, 2008.

\bibitem{villemonteix:2009:iago}
J.~Villemonteix, E.~Vazquez, and E.~Walter.
\newblock An informational approach to the global optimization of
  expensive-to-evaluate functions.
\newblock {\em Journal of Global Optimization}, 44(4):509--534, 2009.

\bibitem{GinsbPhd}
D.~Ginsbourger.
\newblock {\em {M}\'etamod\`eles multiples pour l'approximation et
  l'optimisation de fonctions num\'eriques multivariables}.
\newblock PhD thesis, Ecole nationale sup\'erieure des Mines de Saint-Etienne,
  2009.

\bibitem{dubourg:2011:rbdo}
V.~Dubourg, B.~Sudret, and J.-M. Bourinet.
\newblock Reliability-based design optimization using kriging surrogates and
  subset simulation.
\newblock {\em Structural Multisciplinary Optimization}, 44(5):673--690, 2011.

\bibitem{Dubourg:2011:phd}
V.~Dubourg.
\newblock {\em Adaptive surrogate models for reliability analysis and
  reliability-based design optimization}.
\newblock PhD thesis, Universit\'e Blaise Pascal -- Clermont II, 2011.

\bibitem{delmoral:2006:sequential}
Pierre Del~Moral, Arnaud Doucet, and Ajay Jasra.
\newblock Sequential monte carlo samplers.
\newblock {\em Journal of the Royal Statistical Society: Series B (Statistical
  Methodology)}, 68(3):411--436, 2006.

\bibitem{douc:2008:limit}
R.~Douc and \'E. Moulines.
\newblock Limit theorems for weighted samples with applications to sequential
  monte carlo methods.
\newblock {\em The Annals of Statistics}, 36(5):2344--2376, 2008.

\bibitem{dubourg:mbis}
Vincent Dubourg, Fran{\c c}ois Deheeger, and Bruno Sudret.
\newblock Metamodel-based importance sampling for structural reliability
  analysis.
\newblock {\em Preprint submitted to Probabilistic Engineering Mechanics},
  2011.

\bibitem{dubourg:icasp11}
Vincent Dubourg, Fran{\c c}ois Deheeger, and Bruno Sudret.
\newblock Metamodel-based importance sampling for the simulation of rare
  events.
\newblock In {\em 11th International Conference on Applications of Statistics
  and Probability in Civil Engineering (ICASP 11)}, 2011.

\bibitem{robert:2004:monte}
Christian~P. Robert and G.~Casella.
\newblock {\em Monte Carlo statistical methods, 2nd edition}.
\newblock Springer Verlag, 2004.

\bibitem{santner:2003:dace}
T.~J. Santner, B.~J. Williams, and W.~Notz.
\newblock {\em {T}he {D}esign and {A}nalysis of {C}omputer {E}xperiments}.
\newblock Springer Verlag, 2003.

\bibitem{hol:2006:resampling}
J.~D. Hol, T.~B. Schon, and F.~Gustafsson.
\newblock On resampling algorithms for particle filters.
\newblock In {\em IEEE Workshop on Nonlinear Statistical Signal Processing
  Workshop}, pages 79--82, 2006.

\bibitem{douc2005comparison}
Randal Douc and Olivier Capp{\'{e}}.
\newblock Comparison of resampling schemes for particle filtering.
\newblock In {\em Proceedings of the 4th International Symposium on Image and
  Signal Processing and Analysis (ISPA)}, pages 64--69, 2005.

\bibitem{bolic2003new}
M.~Bolic, Petar~M. Djuri{\'{c}}, and S.~Hong.
\newblock New resampling algorithms for particle filters.
\newblock In {\em IEEE International Conference on Acoustics, Speech, and
  Signal Processing, 2003. Proceedings.(ICASSP'03)}, volume~2, pages 589--592,
  2003.

\bibitem{gayton2003cq}
N.~Gayton, J.M. Bourinet, and M.~Lemaire.
\newblock Cq2rs: a new statistical approach to the response surface method for
  reliability analysis.
\newblock {\em Structural Safety}, 25(1):99--121, 2003.

\bibitem{rajashekhar1993new}
M.R. Rajashekhar and B.R. Ellingwood.
\newblock A new look at the response surface approach for reliability analysis.
\newblock {\em Structural Safety}, 12(3):205--220, 1993.

\bibitem{loeppky:2009:css}
Jason~L. Loeppky, Jerome Sacks, and William~J. Welch.
\newblock Choosing the sample size of a computer experiment: A practical guide.
\newblock {\em Technometrics}, 51(4):366--376, 2009.

\bibitem{Ste99}
M.~L. Stein.
\newblock {\em Interpolation of Spatial Data: Some Theory for {K}riging}.
\newblock Springer, New York, 1999.

\end{thebibliography}
%---------------------------------------------------------------------------------------------------------------------------------------------------------------------
%
%
%
%
\end{document}